\documentclass{elsart1p}
\usepackage{amssymb}
\usepackage{graphicx}
\begin{document}

\begin{frontmatter}
\title{Hadrons in Nuclei: Experiments and Perspectives}
\author{Susan Schadmand}
\address{Institut f\"ur Kernphysik, Forschungszentrum J\"ulich, Germany}

\begin{abstract}

The question of the origin of hadron masses is one major issue in
the understanding of the strong interaction.
The challenge is addressed by searching for indications of in-medium
modifications of hadron properties and studying hadrons in nuclei.
The quest driving in-medium studies is to understand the
origin of hadron masses in the context of spontaneous chiral
symmetry breaking.

The experimental status of the modification of hadron properties in
the nuclear medium is discussed including experiments using hadron,
heavy-ion, and photon beams.
Particular emphasis is put on the production of light mesons from nuclei.

A number of experimental programs is underway to provide a detailed comparison
of properties of free hadrons and hadrons embedded in nuclei.
The existing experimental efforts are discussed and
possibilities are introduced for the new WASA-at-COSY
facility, initially focussed on investigations of symmetries and
symmetry breaking, to contribute to the field.

\end{abstract}


\end{frontmatter}

\section{Introduction}

The fundamental questions concerning the strong interaction and its understanding within
Quantum Chromo Dynamics (QCD) can be related to two basic issues, namely, confinement
and the origin of mass.
One goal of hadron physics is to understand the origin of hadron masses in the context
of spontaneous chiral symmetry breaking.
Chiral symmetry, the fundamental symmetry of QCD for massless quarks, is broken on the
hadron level.
Evidence for this phenomenon are hadrons which are candidates for parity doublets and,
in a chirally symmetric world, should be mass degenerate.
In reality, chiral partners like $\pi$ and $\sigma$ mesons or $\rho$ and $a_1$
mesons bear quite different masses.
In the case of nucleon resonances, an impressive example is the mass difference between
the ground state (938~MeV) and the chiral partner, the S$_{11}$(1535) resonance.

In the field of hadron physics, experimental studies are performed with electromagnetic
and hadronic probes and aim at the investigation of hadron properties and hadronic
interactions.
The study of in-medium properties of mesons and nucleon resonances
carries the exciting promise to find signatures for partial chiral symmetry
restoration at finite baryon density and temperature.

\section{Nucleon Resonances}

Nucleon resonances in the medium are subject to Pauli-blocking of final states
decreasing their width, collision broadening and broadening by the coupling to mesons
with medium-modified properties.
In experiments on total photoabsorption it is found that the second resonance
bump, the excitation region above the $\Delta$(1232), resonance is completely suppressed
for all nuclei from beryllium onward \cite{Bianchi:1994ax}.
This observation is intriguing. However, total photoabsorption does not provide information
about the behavior of individual resonances.
The second resonance region is composed of the overlapping states
P$_{11}$(1440), D$_{13}$(1520), and S$_{11}$(1535).
A significant in-medium modification of the D$_{13}$(1520) is predicted due to the strong
coupling to $N\rho$
while only small effects are expected for the S$_{11}$(1535) state \cite{Post:2004}.
Photoproduction of $\eta$ mesons in this energy range is dominated by the S$_{11}$(1535)
resonance \cite{Krusche:1995nv,Krusche:1997jj,Renard:2000iv,Dugger:2002ft,Crede:2003ax},
while single $\pi^o$, double $\pi^o$ and $\pi^o\pi^{\pm}$ photoproduction
\cite{Harter:1997jq,Wolf:2000qt,Langgartner:2001sg,Bartholomy:2004uz}
show a clear signal for the D$_{13}$(1520) resonance.
Therefore, the reactions are well suited for the study of the in-medium properties
of the respective resonances.

Photoproduction of $\eta$ mesons off nuclei had previously been studied up to
800~MeV incident photon energy with TAPS at MAMI \cite{Roebig_Landau:1996xa} and for
energies up to 1.1 GeV at KEK \cite{Yorita:2000bu} and Tohoku \cite{Kinoshita:2005ca}.
The first experiment did not observe an in-medium broadening of the resonance beyond
effects from Fermi smearing and $\eta$ final state interactions.
The KEK experiment reported some collision broadening of the resonance
and the Tohoku experiment indicated a significant contribution of a higher
lying resonance to the $\gamma n\rightarrow n\eta$ reaction.
However, the experiments do not cover the full line shape of the S$_{11}$(1535).

\begin{figure}[htb]
\begin{center}
\includegraphics[width=0.45\textwidth]{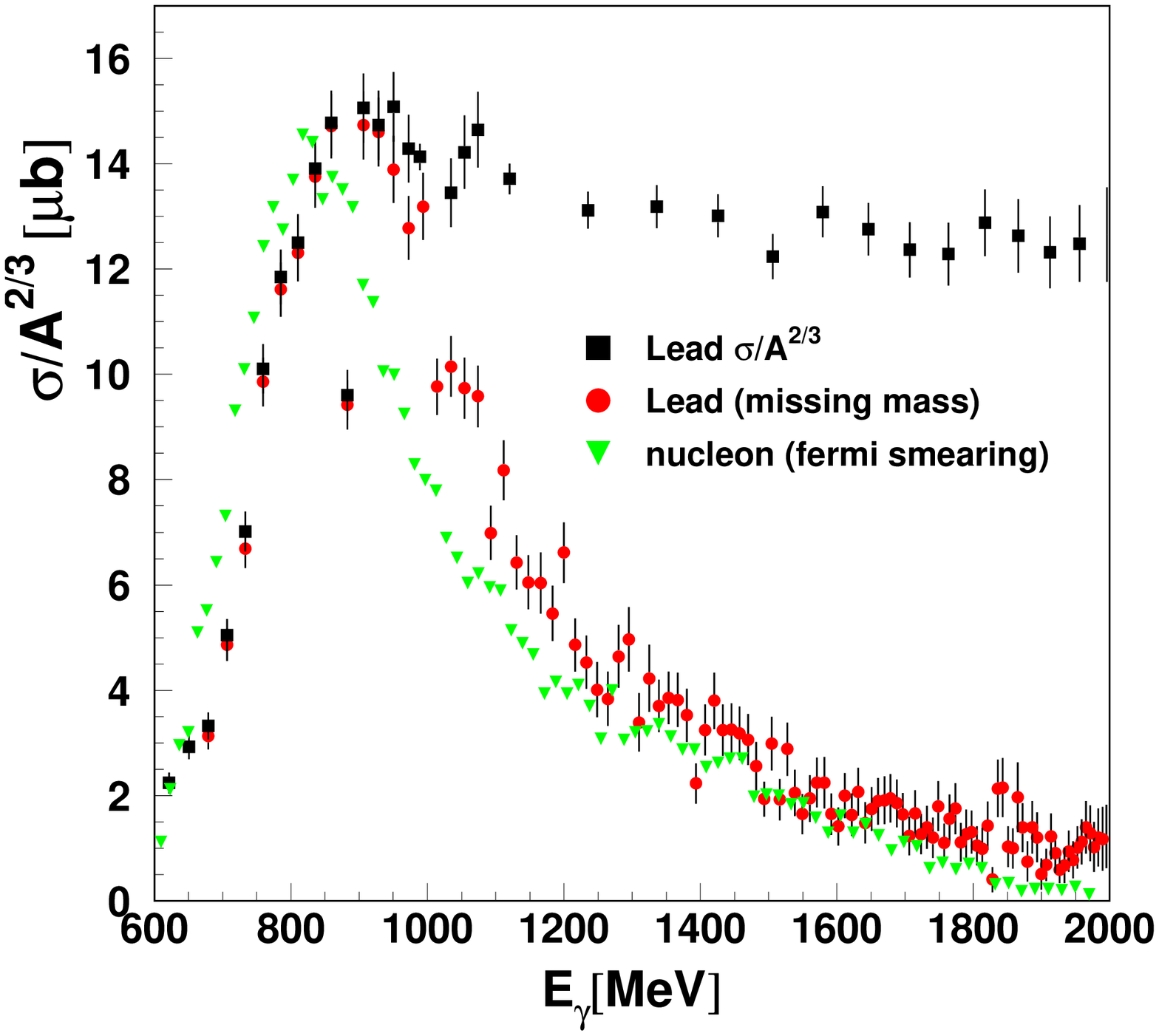}
\hspace*{5mm}
\includegraphics[width=0.4\textwidth]{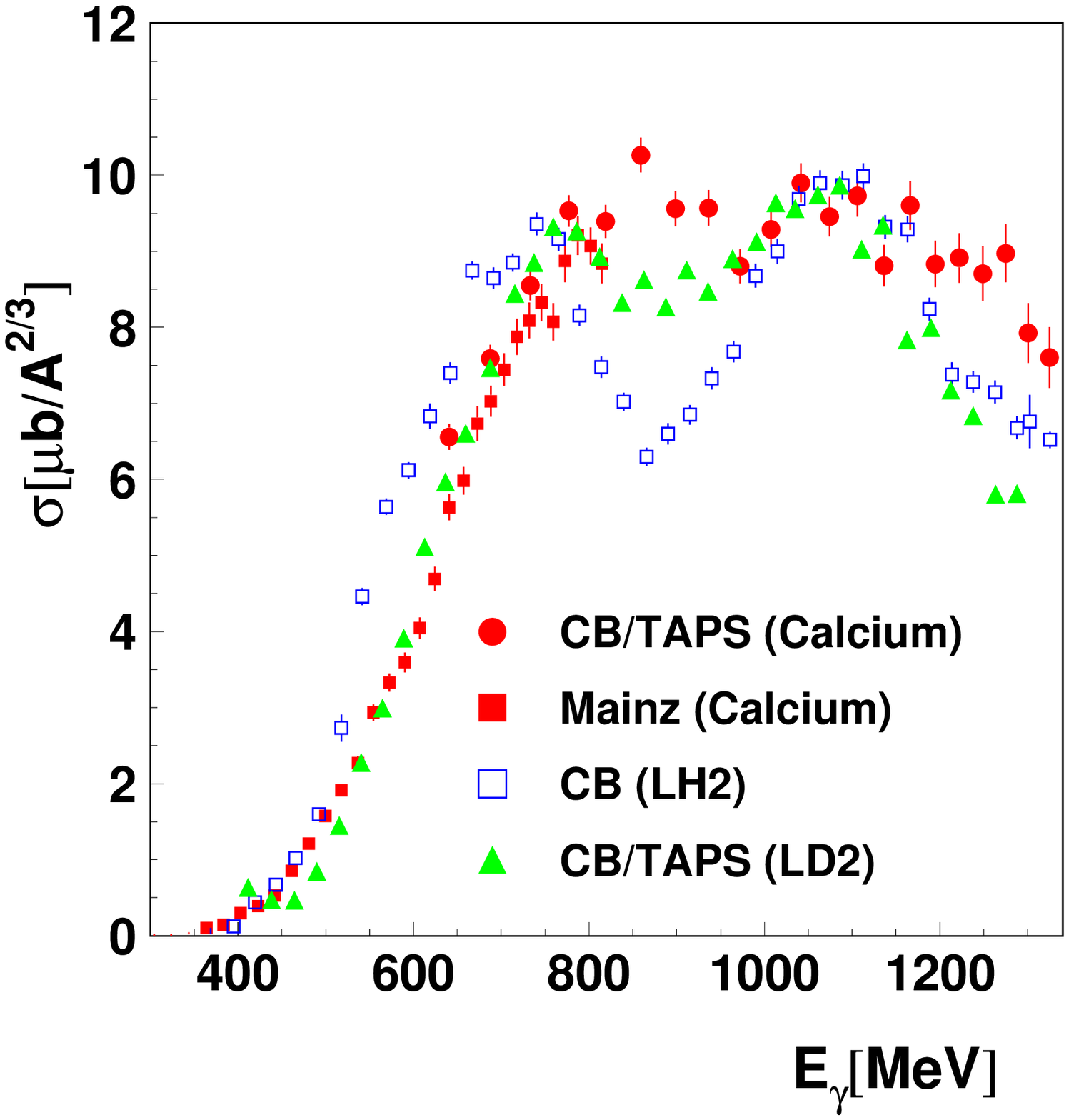}
\end{center}
\caption{
Preliminary results from \cite{Krusche:2006ki}.
Left:
Inclusive cross section and single $\eta$ production cross sections for lead
compared to the Fermi smeared average nucleon cross section.
Right:
Total cross sections for double $\pi^o$ off proton, deuteron
and $^{40}$Ca.
}\label{fig:NucleonResonances}
\end{figure}

Preliminary results for $\eta$ photoproduction from nuclei by the CBELSA/TAPS
experiment are summarized in the left panel of Fig.~\ref{fig:NucleonResonances}.
As already reported in Ref. \cite{Roebig_Landau:1996xa}, the inclusive nuclear
cross sections scale like $A^{2/3}$ for incident photon energies below 800~MeV.
They behave differently at higher incident photon energies where
$\eta\pi$ final states and secondary production mechanisms
(e.g. $\gamma N\rightarrow N\pi$, $\pi N\rightarrow \eta N$)
contribute and obscure the S$_{11}$ line shape.
The contributions can almost completely be suppressed by cuts on the reaction kinematics.
After these cuts, single $\eta$ photoproduction off heavy nuclei, like lead, becomes very
similar to the Fermi smeared average nucleon cross section.
The small discrepancy is at least partly due to inefficiencies of the kinematic cuts.

Preliminary results for double pion photoproduction are summarized in the right panel
of Fig.~\ref{fig:NucleonResonances}.
Results for $\pi^o\pi^o$ and $\pi^o\pi^{\pm}$ have been obtained \cite{Bloch:2007ka}
for incident photon energies up to 800~MeV.
In case of $\pi^o\pi^o$, the excitation functions off the deuteron and off $^{40}$Ca
have exactly the same shape so that there is no indication for an in-medium modification
of the D$_{13}$(1520) at energies below its peak position.
However, the model of \cite{Post:2004} predicts the main effect for the high energy tail of
the resonance.
This part is covered by the preliminary data from the CBELSA/TAPS experiment
(see Fig.~\ref{fig:NucleonResonances}, right hand side).
Here, a difference between deuteron and nuclear data is seen, in particular in the
region of the valley between the second and third resonance bump.

A significant in-medium modification of the S$_{11}$(1535) line shape is not observed.
For the D$_{13}$(1520) resonance, still under analysis, a slight suppression seems possible.
Further investigations could concentrate on the comparison with $\pi^o\pi^{\pm}$ and
$\pi^+\pi^-$ pion pairs which can stem from an in-medium $\rho$ meson while the decay of
$\rho$ into $\pi^o\pi^o$ is forbidden.

\section{Scalar-Isoscalar Pion Pairs}

A particularly interesting case is the mass split between
the $J^{\pi}$=0$^{-}$ pion and the $J^{\pi}$=0$^{+}$ $\sigma$-meson.
The naive assumption that the two masses should become degenerate
in the chiral limit is supported by model calculations
\cite{Bernard:1987im}.
However, the very nature of the $\sigma$ meson is a matter of debate.
The review of particle properties \cite{Yao:2006px} lists
the $f_{0}$(600) with a mass range from 400 - 1200~MeV and a
full width between 600~MeV and 1000~MeV.
Recently, predictions for mass and width from
dispersion relations have been derived in \cite{Caprini:2005zr}.
The $\sigma$ meson is treated as a pure $q\bar{q}$ (quasi)bound state
\cite{Bernard:1987im,Hatsuda:1999kd,Aouissat:1999vx},
and as a correlated $\pi\pi$ pair in a $I=0$, $J^{\pi}=0^+$
state \cite{Chiang:1997di,Roca:2002vd,Chanfray:2004vb}.
In all cases, a strong coupling to scalar-isoscalar pion pairs and
a significant in-medium modification of the invariant mass
distribution of the pion pairs is predicted.
This is either due to the in-medium spectral function of the $\sigma$ meson
\cite{Hatsuda:1999kd} or the in-medium modification of the pion-pion interaction
\cite{Roca:2002vd} due to coupling to nucleon - hole, $\Delta$ - hole and
$N^{\star}$ - hole states.
The predicted effect is a downward shift of the strength in the invariant
mass distributions of scalar, isoscalar pion pairs in nuclear matter.

First experimental evidence had been reported by the CHAOS collaboration
from the measurement of pion induced double pion production reactions
\cite{Bonutti:1996ij,Bonutti:1999,Bonutti:2000bv,Camerini:2004sz,Grion:2005hu}.
The main finding was a buildup of strength with rising mass number at low
invariant masses for the $\pi^+\pi^-$ final state.
The effect was not observed for the $\pi^+\pi^+$ channel where the $\sigma$
meson cannot contribute.
A similar effect was found by the Crystal Ball collaboration at BNL.
Here, an enhancement of strength at low masses was observed for heavy nuclei
in the $\pi^- A\rightarrow A\pi^o\pi^o$ reaction \cite{Starostin:2000cb}.
In photon induced reactions, pions can be produced in the entire volume
of the nuclei but final state interactions suppress the contributions
from the deep interior of the nuclei.
Final state interactions can be minimized by the use of low incident
photon energies, giving rise to low energy pions which have much
larger mean free paths than pions that can excite
the $\Delta$-resonance \cite{Krusche:2004uw}.
Photoproduction of the different charge states of pions from the free proton
and the quasifree neutron has previously been studied in detail with the DAPHNE
\cite{Braghieri:1995rf,Zabrodin:1997xd,Zabrodin:1999sq,Ahrens:2003na,Ahrens:2005ia}
and TAPS detectors
\cite{Harter:1997jq,Krusche:1999tv,Wolf:2000qt,Kleber:2000qs,Langgartner:2001sg,Kotulla:2003cx}
at MAMI-B in Mainz from threshold to the second resonance region,
and for the $\pi^o\pi^o$ channel at higher incident photon energies at
GRAAL in Grenoble \cite{Assafiri:2003mv}.
See Ref.~\cite{Krusche:2003ik} for an overview.

First results from a measurement of $\pi^o\pi^o$ and $\pi^o\pi^{\pm}$
photoproduction off carbon and lead have been reported in
\cite{Messchendorp:2002au}.
A shift of the strength to lower invariant masses was found for the heavier
nucleus for the $\pi^o\pi^o$ channel but not for the mixed charge channel.
In Ref.~\cite{Bloch:2007ka}, more experimental detail and the results of
an additional measurement of double pion photoproduction off calcium nuclei
are presented and compared to model calculations.
The invariant mass spectra show a similar effect as already reported in
Ref.~\cite{Messchendorp:2002au} for carbon and lead nuclei, namely a softening
of the $\pi^o\pi^o$ distributions relative to the $\pi^o\pi^{\pm}$
distributions.
The strength of the effect is comparable to that from carbon.
The data have been compared to calculations in the framework of the BUU model
\cite{Buss:2006vh}.
A sizable part of the in-medium effects can be explained by the model by final
state interaction effects which tend to shift rescattered pions
to smaller kinetic energies.
Only for the lowest incident photon energies a small additional downward shift
of the strength to small invariant masses for the $\pi^o\pi^o$ channel
may be visible.

Decisive results will come from a recently completed experimental run
\cite{prop-pipi-nucs} with the 4$\pi$ detector combination Crystal Ball
and TAPS at MAMI-B.
Here, superior statistics has been accumulated for carbon, calcium,
and lead targets.

\section{Vector Mesons}

In-medium modifications of vector mesons have been searched for via the
spectroscopy of dilepton pairs in heavy ion reactions by the CERES
experiment \cite{Agakishiev:1995xb,Adamova:2002kf} and more recently the NA60
collaboration \cite{Damjanovic:2005ni} which reported an in-medium broadening
of the $\rho$ meson.
In-medium modifications of $\rho$ and $\omega$ mesons were reported
from 12 GeV p+A reactions at KEK \cite{Naruki:2005kd}.
Here, the authors find that the mass spectra are well reproduced by a model
that takes into account the density dependence of the vector meson mass modification.
In contradiction to these results, the photon induced dilepton experiment g7 with CLAS
at Jefferson Lab concludes that the masses are consistent with the PDG values
and the widths are consistent with the collisional broadening \cite{Djalali:2007qv}.
The results do not show a doubling of the $\rho$ width reported by NA60 and do not
favor the predicted mass shifts of 16-20\%.
This controversy will have to be addressed in the near future.
Finally, a modification of the $\Phi$ meson has been suggested on the basis
of the $A$-dependence of the photoproduction yields \cite{Ishikawa:2004id}.

\begin{figure}[htb]
\begin{center}
\includegraphics[width=0.4\textwidth]{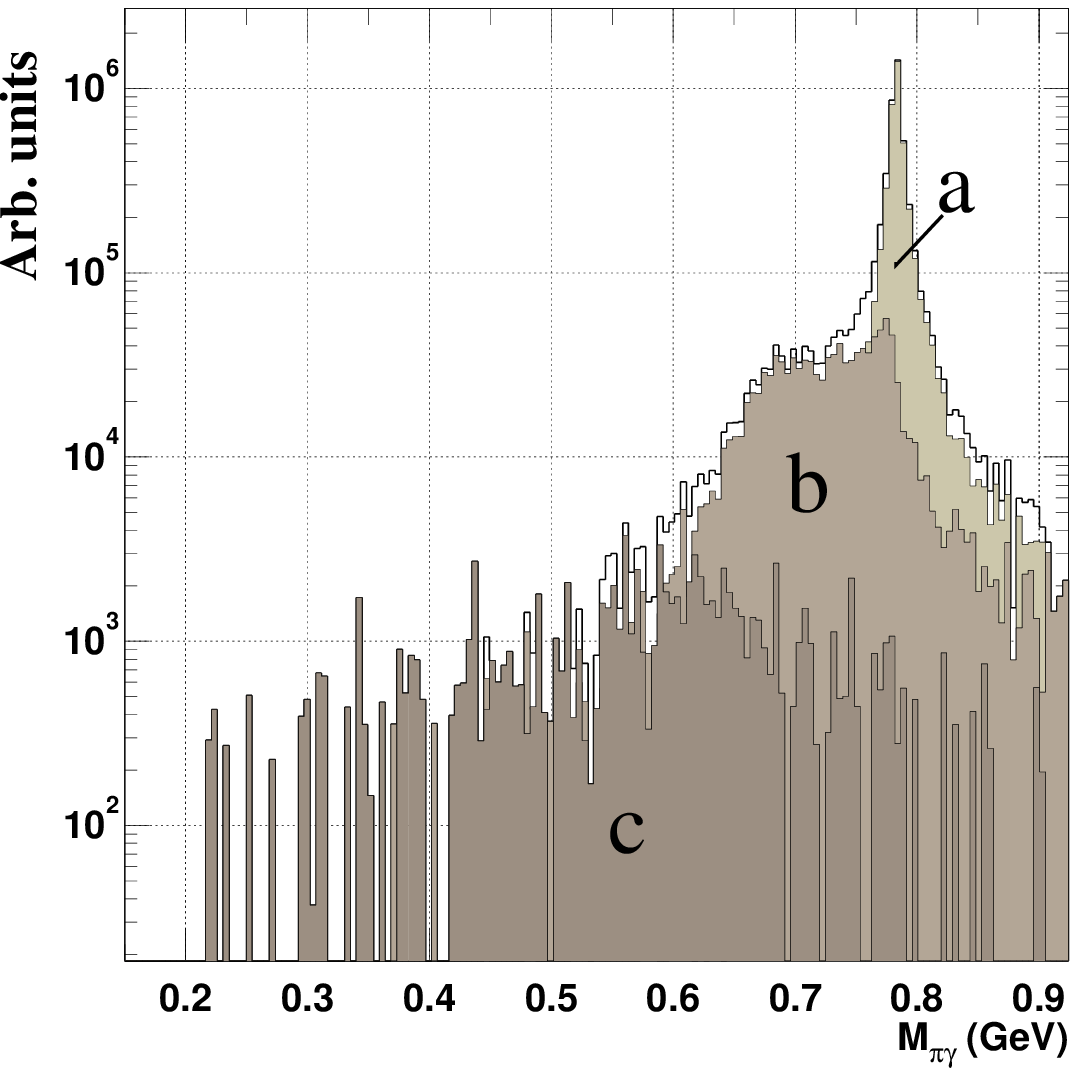}
\hspace*{5mm}
\includegraphics[width=0.5\textwidth]{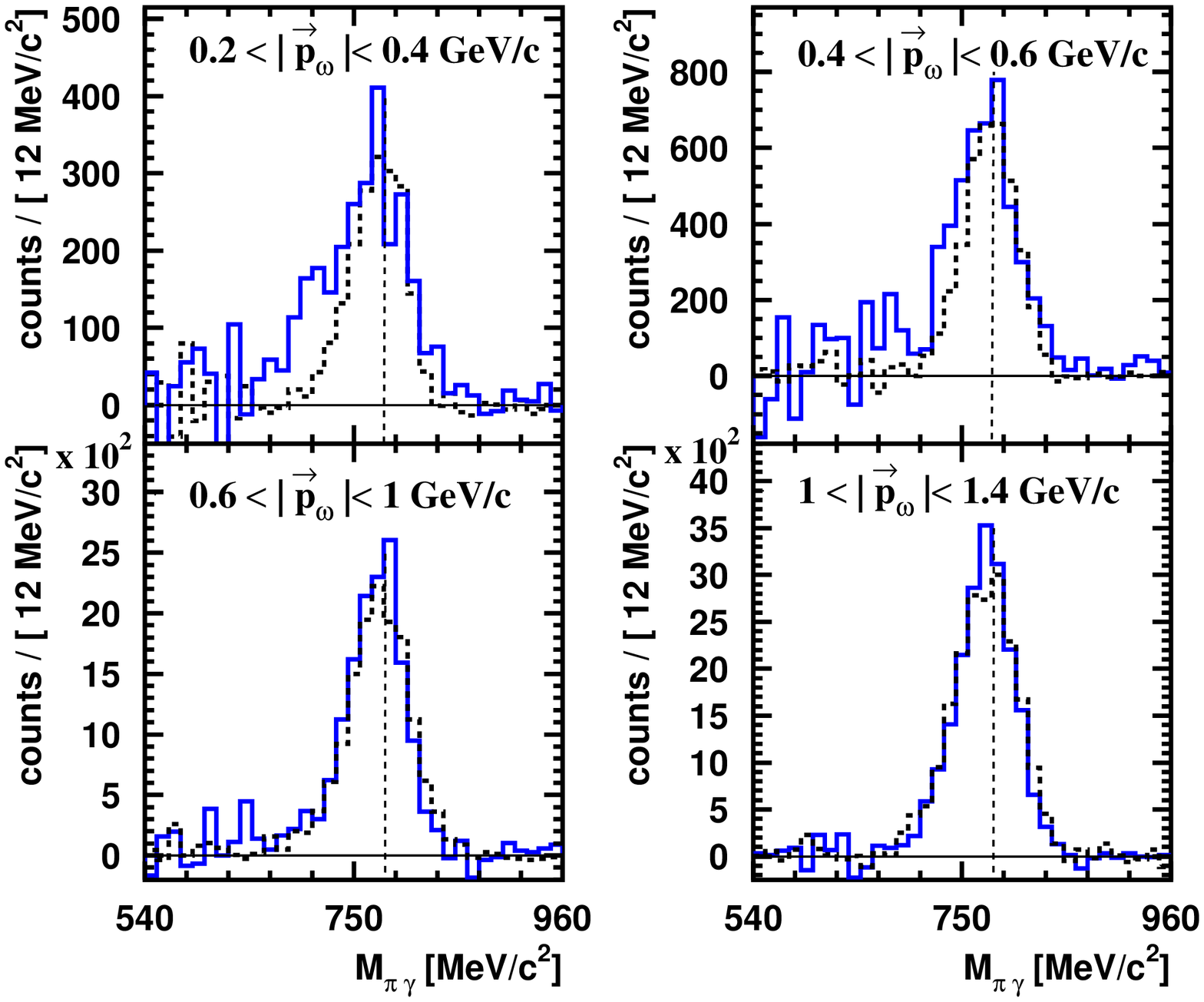}
\end{center}
\caption{
\underline{Left:}
The $\pi^0\gamma$-mass distribution obtained from a Monte-Carlo
simulation of the process $\gamma+Nb\rightarrow \pi^0\gamma+X$ at
$E_{\gamma}$=1.2~GeV. The spectrum is decomposed into different
contributions corresponding to the fraction of $\omega$ mesons decaying
outside ($\rho<0.05$~fm$^{-3}$) the nucleus ({\bf a}),
the fraction of $\omega$ mesons decaying inside ($\rho>0.05$~fm$^{-3}$)
for which the $\pi^0$ does not rescatter ({\bf b}), and the fraction of
$\omega$ mesons decaying inside the nucleus for which $\pi^0$
rescatters ({\bf c}). All with a condition of $T_{\pi^0}>150$~MeV.
From \protect\cite{Messchendorp:2001pa}.
\underline{Right:}
$\pi^0 \gamma$ mass spectrum after background subtraction and FSI suppression
($T_{\pi^0} > 150$ ~MeV) for different $\omega$ momentum bins.
Solid histogram: Nb data, dashed histogram: $\rm{LH_2}$ data.
From \protect\cite{Trnka:2005ey}.
}
\label{fig:prl-94-192303-fig3}
\end{figure}

The CBELSA/TAPS \cite{Trnka:2005ey} experiment measured the line shape of the
$\omega\rightarrow\pi^o\gamma$ invariant mass peak from the free proton and from
nuclei.
The measured invariant mass peaks of the $\pi^o\rightarrow\gamma\gamma$,
$\eta\rightarrow 3\pi^o\rightarrow 6\gamma$, and
$\eta'\rightarrow\pi^o\pi^o\eta\rightarrow 6\gamma$ decays were identical
for hydrogen and the nuclear targets.
Only in case of the $\omega$, a low energy shoulder of the peak was found
for nuclei and is shown in the right panel of Fig. \ref{fig:prl-94-192303-fig3}.
Due to its life time, only a small fraction of
the $\omega$ mesons decay in the medium, so that also the nuclear invariant mass
peaks include a dominant contribution of unmodified in-vacuum decays.
This was anticipated beforehand and simulated as shown in the left panel of
Fig. \ref{fig:prl-94-192303-fig3}.
Indeed, in the data the low-energy tail of the invariant mass
stems almost exclusively from $\omega$ mesons with small momenta, which have the
largest change for in-medium decays.
Similar results where found for carbon, an analysis of the $A$ scaling of the
production cross sections is under way, and a second generation experiment is approved.
A detailed report of the most recent developments is given by M.~Kotulla
(Giessen University) at this conference, INPC2007.

\section{Summary and Outlook}

With the variety of results discussed above it remains obvious that further studies are
necessary and of high interest.
The possible reasons for medium effects on hadrons range from trivial
to enticing causes:

\begin{itemize}
\item absorption and rescattering of mesons
\item modified hadron-hadron interaction
\item partial chiral symmetry restoration
\item meson-baryon coupling
\item meson-nucleus attractive potential
\item mass shift
\item broadening
\item bound states (not discussed here)
\end{itemize}

Experiments are in accordance with theoretical scenarios for
changes of hadron properties in the nuclear medium.
However, some controversy to be resolved and the influence of the various effects
has to be quantified.
It is important to know that studying the in-medium behavior of hadrons is a promising
approach to learn more about the origin of their mass.

As a further perspective, I would like to mention the possibility to study
vector mesons in p+A reactions with the newly implemented WASA-at-COSY facility
at the Research center J\"ulich (Germany).
The study of the $\rho$/$\omega$ line shapes in the nuclear medium
and of $\phi$ mesons in medium could be performed via
dilepton production (elementary reactions: J.~Stepaniak et al. in \cite{Adam:2004ch}).
Furthermore, one could envisage the simultaneous measurement of the Dalitz decay
$\pi^o\gamma$ of the $\omega$ meson and a comparison to photon induced reactions,
elementary and heavy ion dilepton production.
This procedure could provide valuable information on the final state interactions
of pions in medium.
The study of the $\omega\to\pi^o\gamma$ channel in p+A reactions was suggested
in \cite{Sibirtsev:2000th} and later for photon induced reactions in
\cite{Messchendorp:2001pa}.
WASA-at-COSY \cite{Adam:2004ch}
is a 4$\pi$ detection system that can detect neutral and charged decays,
even dileptons, and that can handle high rates.
The venture would become possible in second generation WASA-at-COSY experiments
that can employ nuclear targets and would be very much suited for studies of
meson production and decays in the nuclear medium.

\end{document}